\documentclass[a4paper,11pt]{article}
\usepackage{todonotes}
\usepackage{geometry}
\usepackage{xspace}
\usepackage{threeparttable}
\usepackage{subcaption} 
\usepackage{listings}

\newcommand{\progname}{\textsf}
\newcommand{\compname}{\emph} 
\newcommand{\compswitch}{\texttt}

\newcommand{\lrs}{\progname{lrs}\xspace}

\newcommand{\mplrs}{\progname{mplrs}\xspace}
\newcommand{\mplrsgmp}{\progname{mplrsgmp}\xspace}
\newcommand{\lrsa}{\progname{lrs1}\xspace}

\newcommand{\lrsb}{\progname{lrs2}\xspace}

\newcommand{\lrsgmp}{\progname{lrsgmp}\xspace}
\newcommand{\lrsflint}{\progname{lrsflint}\xspace}
\newcommand{\lrsmp}{\progname{lrs\_mp}\xspace}
\newcommand{\lrsMP}{\progname{lrsMP}\xspace}
\newcommand{\coll}{\progname{coll}\xspace}
\newcommand{\colla}{\progname{coll1}\xspace}
\newcommand{\collan}{\progname{coll1n}\xspace}
\newcommand{\collb}{\progname{coll2}\xspace}
\newcommand{\collbn}{\progname{coll2n}\xspace}
\newcommand{\collgmp}{\progname{collgmp}\xspace}
\newcommand{\collflint}{\progname{collflint}\xspace}
\newcommand{\collmp}{\progname{collmp}\xspace}

\newcommand{\fixed}{\progname{fixed.c}\xspace}
\newcommand{\fixedo}{\progname{fixed1}\xspace}
\newcommand{\fixedon}{\progname{fixed1n}\xspace}
\newcommand{\fixedt}{\progname{fixed2}\xspace}
\newcommand{\fixedtn}{\progname{fixed2n}\xspace}
\newcommand{\fixedg}{\progname{fixedgmp}\xspace}
\newcommand{\fixedm}{\progname{fixedmp}\xspace}

\newcommand{\hybrid}{\progname{hybrid}\xspace}

\newcommand{\make}{\progname{makefile}\xspace}
\newcommand{\lrslong}{\progname{lrslong}\xspace}
\newcommand{\lrslib}{\progname{lrslib}\xspace}
\newcommand{\lmp}{\progname{lrs\_mp}\xspace}
\newcommand{\runo}{\progname{run1}\xspace}
\newcommand{\runt}{\progname{run2}\xspace}
\newcommand{\runmp}{\progname{runmp}\xspace}

\newcommand{\run}{\progname{run}\xspace}
\newcommand{\main}{\progname{main}\xspace}
\newcommand{\setjmp}{\progname{setjmp}\xspace}
\newcommand{\lov}{\progname{lrs\_overflow}\xspace}
\newcommand{\Mm}{\progname{MAXDm}\xspace}
\newcommand{\Ml}{\progname{MAXDl}\xspace}
\newcommand{\Ma}{\progname{MAXDa}\xspace}

\newcommand{\redund}{\progname{redund}\xspace}
\newcommand{\bufo}{\progname{buf1}\xspace}
\newcommand{\suf}{\progname{suf}\xspace}
\newcommand{\lrsarith}{\progname{lrsarith}\xspace}

\newcommand{\mplrsflint}{\progname{mplrsflint}\xspace}
\newcommand{\mplrsa}{\progname{mplrs1}\xspace}

\newcommand{\mplrsb}{\progname{mplrs2}\xspace}

\newcommand{\normaliz}{\progname{normaliz}\xspace}

\newcommand{\maif}{\compname{mai48}\xspace}

\newcommand{\polytope}{\emph}
\newcommand{\bvseven}{\polytope{bv7}\xspace}
\newcommand{\cthirty}{\polytope{c30}\xspace}
\newcommand{\cforty}{\polytope{c40}\xspace}
\newcommand{\fqfour}{\polytope{fq48}\xspace}

\newcommand{\mitseven}{\polytope{mit71}\xspace}
\newcommand{\mitine}{\polytope{mit}\xspace}
\newcommand{\permten}{\polytope{perm10}\xspace}

\newcommand{\cpseven}{\polytope{cp7}\xspace}
\newcommand{\peight}{\polytope{p8-6}\xspace}
\newcommand{\vffive}{\polytope{vf500}\xspace}
\newcommand{\kmtwo}{\polytope{km22}\xspace}
\newcommand{\vfnine}{\polytope{vf900}\xspace}

\newcommand{\cmax}{\mathit{max}}

\usepackage{amsmath}
\usepackage{amssymb}
\usepackage{latexsym}
\usepackage{graphicx}
\usepackage{color}
\usepackage{hyperref}
\usepackage{slashbox}
\usepackage{algorithm}
\usepackage{algpseudocode}

\definecolor{darkblue}{rgb}{0,0,0.6}
\hypersetup{
	citecolor=darkblue,
	linkcolor=darkblue,
	urlcolor=darkblue,
	breaklinks=true,
	colorlinks=true,
	citebordercolor=0 0 0,
	filebordercolor=0 0 0,
	linkbordercolor=0 0 0,
	menubordercolor=0 0 0,
	urlbordercolor=0 0 0,
	pdfhighlight=/I,
	pdfborder=0 0 0,
	bookmarksopen=true,
	bookmarksnumbered=true
}

\include{./mathlib}
\include{./macros}

\begin{document}

\title{\lrsarith: a small fixed/hybrid arithmetic C library}
\author{David Avis \\ School of Informatics, Kyoto University, Kyoto, Japan and
   \\ School of Computer Science,
    McGill University, Montr{\'e}al, Qu{\'e}bec, Canada
\and Charles Jordan \\
 Graduate School of Information Science and Technology,
Hokkaido University, Sapporo, Japan}
\maketitle
\begin{abstract}
We describe \lrsarith which is a small fixed precision and 
hybrid arithmetic C library for integers and
rationals that we developed for use in the
\lrslib library for polyhedral computation. 
Using a generic set of operations, a program can be compiled with either 64-bit
or 128-bit (if available) fixed precision, with an extended precision library 
such as GMP or the built-in MP routines. A simple scheme checks for overflow
and either terminates the program or, in hybrid mode, changes to a higher precision arithmetic.
Implementing these arithmetics in \lrslib resulted in only minimal changes to the original code.
We give computational results using \lrs and \mplrs, vertex/facet enumeration codes in \lrslib, using 64 and 128 bit fixed integer arithmetic with and without overflow checking, 
GMP arithmetic, \lrsarith hybrid arithmetic with both GMP and MP, and 
FLINT hybrid arithmetic.
We give a small self-contained example C program using the \lrsarith package in both
fixed precision and  hybrid mode.
\end{abstract}

\section{Introduction}
When writing mathematical software, a fundamental choice is the arithmetic
to use.
It is easy to be seduced by the performance of native integers, but the
possibility of incorrect output caused by overflow can be a major concern.
To guarantee correctness, in many cases one must rely on multiprecision
arithmetic such as that provided by the GNU Multiple Precision (GMP) Arithmetic
Library.  This comes with
a large performance penalty compared to native integers on inputs
where they suffice, and it's tempting to maintain multiple versions with
different arithmetic.

Until version 7, this was the situation with \lrslib, a library of programs
for polyhedral computation including e.g. 
\lrs for vertex/facet enumeration and \redund for redundancy removal.
From the outset \lrslib used exact arithmetic and could be compiled with either
fixed precision C integer arithmetic (without overflow checking)
or with an internal extended precision arithmetic library.
For safety, the default was to use extended precision.
%
Later versions could also be compiled with the GMP
package and more recently, with the hybrid arithmetic package available
in FLINT.  The fixed precision versions perform several times faster than
the extended precision versions but can produce incorrect results if
overflow occurs.

For this reason we developed a simple scheme to detect the possibility of
overflow occurring while using fixed precision arithmetic.
The code could then halt and allow the user to restart using
higher or extended precision arithmetic.
This led to the development of a hybrid scheme in \lrslib to allow
automatic restart  with the next higher precision arithmetic.  This gives
users the performance of native arithmetic in most cases where it is safe,
while keeping the correctness provided by the extended precision versions.
The restart is from (very near) the point where a possible overflow was
detected.
Implementing this in \lrslib resulted in only minimal changes to the original
code and gives a significant performance improvement.

The main purpose of this paper is to explain how overflow checking and
hybrid arithmetic was implemented and to give numerical results for a case study
involving vertex/facet enumeration.  This explains the improved performance
of \lrslib 7 compared to previous versions.
These techniques can be used in any C code without requiring the full
\lrslib library, so we created the much smaller, independent \lrsarith
package which contains only the arithmetic code. It can
be downloaded from \cite{lrs}.

Implementing arithmetic in \lrsarith allows developers to postpone the
choice of arithmetic until compile time.  With some additional effort
to handle overflows and restarting from intermediate points of the
computation, it also allows the developer
to obtain the performance of native integers when they
suffice and the correctness of an extended precision library when required.

\subsection{Related work}
\label{subsec:relwork}

When comparing various codes for vertex/facet enumeration~\cite{AJ15b},
we noted that the hybrid arithmetic available in \normaliz could give
a large speedup for certain inputs.  The default in \normaliz is to
begin the computation in 64-bit integers and to restart using GMP
extended precision if overflow occurs.  However, the restart is from
the beginning of the computation, which can hurt performance on
certain inputs where the repeated work is significant (see the
comparison in~\cite{AJ15b}).

The Fast Library for Number Theory (FLINT)~\cite{flint} is a collection
of modules that support various arithmetic operations.  In particular,
we use the \texttt{fmpz} module for extended precision arithmetic that
balances ``very efficient'' performance on small integers with
performance similar to GMP for large integers.  We will compare
\lrsarith performance with FLINT in Section~\ref{subsec:arith}.  
Although FLINT is easier to develop with since switching 
arithmetic is transparent to the user, \lrsarith can give
higher performance on small integers. FLINT contains
much additional functionality beyond extended precision arithmetic,
but usually requires the user to install it.

Porta~\cite{porta} is a collection of routines related to polytopes, including
vertex/facet enumeration.  It supports both fixed precision and its
own multiprecision arithmetic and was included in the comparison~\cite{AJ15b}
mentioned earlier.

The standard general purpose library for extended precision arithmetic
with large integers is GMP\footnote{\url{https://gmplib.org/}}.  It is
highly optimized for many platforms and was the default in \lrslib for
some time.  Historically, \lrslib used its own multiprecision 
arithmetic (MP) and this is still supported.  Comparisons are given
in Section~\ref{subsec:arith}.

\subsection{Organization of the paper}
\label{subsec:organ}

The paper contains two parts that can essentially be read independently.
The first part gives a detailed explanation of how we implement the various
arithmetics with a simple but complete example program that can be used as a 
template for developing with \lrsarith.
In Section~\ref{sec:fixed} we show how overflow checking is implemented in our fixed
precision \lrslong library. We give a simple example to show how a 
single C code can be compiled with 64-bit, 128-bit,
MP or GMP arithmetic. In the first three cases the program terminates when the 
possibility of overflow is detected. 
In Section \ref{sec:hybrida} we continue the example to show how a program can 
restart after overflow has been detected, using the next level of arithmetic.
This form of hybrid arithmetic is used in \lrslib.
Computational results are presented in both sections for a larger example involving
Collatz sequences.

The second part of the paper begins in
Section \ref{sec:ve} where we present our case study and give
computational results comparing various arithmetic packages that can be used in \lrslib.
These include fixed integer arithmetic with and without overflow checking,
GMP arithmetic, two versions of \lrsarith hybrid arithmetic and
FLINT hybrid arithmetic.

\section{Fixed arithmetic}
\label{sec:fixed}
\subsection{Definitions and overflow handling}
Arithmetic is handled in \lrsarith by defining a generic integer type and a set of
generic operations. A generic integer \texttt{a}, integer vector \texttt{v}
and integer matrix \texttt{A} are defined
\begin{verbatim} 
lrs_mp a;     lrs_mp_vector v;    lrs_mp_matrix A;
\end{verbatim} 
allocated
\begin{verbatim} 
lrs_alloc_mp(a);     v=lrs_alloc_mp_vector(n);    A=lrs_alloc_mp_matrix(m,n);
\end{verbatim} 
and freed
\begin{verbatim} 
lrs_clear_mp(a);     lrs_clear_mp_vector(n);    lrs_clear_mp_matrix(m,n);
\end{verbatim} 
where \texttt{m} and \texttt{n} are the dimensions. The types are assigned 
at compile time depending on the arithmetic used.
For 64-bit and 128-bit integers they are assigned to fixed length integers 
and for GMP arithmetic to the GMP integer type:
\begin{verbatim}
typedef long long lrs_mp[1];                /*  one long integer     */
typedef __int128 lrs_mp[1];                 /*  one 128-bit integer  */
typedef long long lrs_mp[MAX_DIGITS + 1];   /*  one MP integer       */
typedef mpz_t lrs_mp;                       /*  one GMP integer      */
\end{verbatim}

Operations using \lmp integers are written generically. Typical examples are
as follows where \texttt{a},\texttt{b},\texttt{c},\texttt{d},\texttt{e} are 
\lmp integers, \texttt{i} is a \texttt{long long} and the equivalent C code
is given in parentheses:
\begin{verbatim}
itomp(i,a);           (a=i)
copy(b,a);            (b=a)     
addint(a,b,c);        (c=a+b)
mulint(a,b,c);        (c=a*b)
divint(a,b,c);        (c=a/b a=a%b)
linint(a, ka, b, kb); (a=a*ka+b*kb for ka, kb long long integers)
qpiv(a,b,c,d,e);      (a=(a*b-c*d)/e  where the division is exact )
\end{verbatim}
A small C program, \fixed, using some of these operations is given in Appendix \ref{app:fixed}.
It reads in an integer $k$ and attempts to repeatedly square it six times.
We will discuss the program later in this section.
Generic operations are either assigned at compile time by macros, for example:
\begin{verbatim}
#define addint(a,b,c)         *(c) = *(a) + *(b)          /* 64-bit or 128-bit */
#define addint(a,b,c)         mpz_add((c),(a),(b))        /* GMP               */
\end{verbatim}
or by C code in the case of MP.
In this way arithmetic operations are essentially the same as using the underlying arithmetic
package. 

The problem is that overflow is not detected in 64-bit and 128-bit arithmetic.
We solve this problem by a technique we call {\em lazy overflow handling}. Using very
light extra computation we detect when it is {\em possible} that an overflow may
occur. Control is then given to an overflow handler that either halts computation or
restarts the program using arithmetic with higher precision. 
While restarting from the beginning of execution is simplest, later we will see
options for restarting from intermediate checkpoints -- allowing programs to
use faster arithmetic for initial portions of computations in some cases.

To incorporate lazy overflow handling we define some constants that depend on the word size
$W$ which is either 64 or 128 in \lrsarith:
\begin{equation}
 \Mm = \left\lfloor \sqrt{2^{W-1}-1} \right\rfloor~~~ 
 \Ml = 2^{W/2-1}-1 ~~~ \Ma = 2^{W-2}-1
\end{equation}
$\Ma$ is the constant used in testing for overflow when using addition or 
similar operations, $\Ml$ is used for \texttt{linint} and $\Mm$ is used for
multiplying.  Three macros test whether the operands \texttt{a} 
and \texttt{b} are out of bounds:
\begin{verbatim}
#define mpsafem(a,b)          *(a)> MAXDm ||*(b)> MAXDm ||*(a)<- MAXDm ||*(b)<- MAXDm 
#define mpsafel(a,b)          *(a)> MAXDl ||*(b)> MAXDl ||*(a)<- MAXDl ||*(b)<- MAXDl 
#define mpsafea(a,b)          *(a)> MAXDa ||*(b)> MAXDa ||*(a)<- MAXDa ||*(b)<- MAXDa 
\end{verbatim}
Using these macros we can write  macros for the generic operations above as follows:
\begin{verbatim}
#define addint(a,b,c)      {if(mpsafea(a,b)) lrs_overflow(1); else *(c) = *(a) + *(b);}
#define mulint(a,b,c)      {if(mpsafem(a,b)) lrs_overflow(1); else *(c) = *(a) * *(b);}
#define divint(a,b,c)      {*(c) = *(a) / *(b); *(a) = *(a) % *(b);}
#define linint(a,ka,b,kb)  {if(mpsafel(a,b)) lrs_overflow(1); else *(a)=*(a)*ka+*(b)*kb;}
#define qpiv(a,b,c,d,e)    {if(mpsafel(a,b)|| mpsafel(c,d)) lrs_overflow(1);  
                           else *(a) =(*(a) * *(b) - *(c) * *(d))/(*e);}
\end{verbatim}
We claim that if the integer arithmetic would overflow then \lov is called. 
Since we are using signed integer arithmetic this is equivalent to proving that
the result (and intermediate values) in each case is at most $2^{W-1}-1$.
Proceeding case by case:
\begin{itemize}
\item
\texttt{addint}($a,b,c$): $|a|,|b| \le  \Ma$ and so $a+b  \le  2*(2^{W-2}-1) = 2^{W-1}-2$\,;
\item
\texttt{mulint}($a,b,c$): $|a|,|b| \le  \Mm$ and so $a*b  \le  \left(\sqrt{2^{W-1}-1}\right)^2 = 2^{W-1}-1$\,;
\item
\texttt{linint}($a, ka, b, kb$)  $|a|,|b|,|ka|,|kb| \le \Ml$ and so 

$~~~~~a*ka+b*kb \le 2\,(2^{W/2-1}-1)^2 = 2^{W-1}-2^{W/2+1}+2$\,.
\end{itemize}
Note that \texttt{divint} cannot overflow and that the analysis for 
\texttt{qpiv} is thus essentially the same as for \texttt{linint},
proving the claim.

Rational arithmetic is handled in \lrsarith in a very simple way
by representing the rational number
$a/b$ by two \lrsmp integers \texttt{a} and \texttt{b}. The operation
\begin{verbatim}
reduce(a,b)        
\end{verbatim}
divides \texttt{a} and \texttt{b} by their greatest common divisor.
Arithmetic operations for rationals
are based on the integer operations above. For example
\begin{verbatim}
mulrat(a,b,c,d,e,f)        (e/f = a/b * c/d, with e/f reduced)
\end{verbatim}
is implemented by
\begin{verbatim}
mulint (a,c,e);
mulint (b,d,f);
reduce (e,f);
\end{verbatim}
Overflow checking for rational arithmetic is inherited from 
the corresponding integer
arithmetic.

We now look more closely at the code \fixed for iterated squaring given in
Listing~1 of Appendix~\ref{app:fixed}.
Among the includes on lines 1--3 we have the \lrsarith header file.
On lines 7--9
is the function \lov that handles overflow processing. In this case
it simply prints a message and halts the program. In the main routine we see that
two \lrsmp variables are declared. These are allocated and cleared in lines
16 and 26. These are null operations for 64 or 128-bit arithmetic and MP
but are needed in GMP.
On line 15 we initialize the arithmetic package.
The next lines read an integer $k$ and attempt to iteratively square it six
times.

Using the \make in Listing 2 we can compile \fixed with each of the arithmetic 
packages depending on the compiler switches used.
For 64 or 128-bit arithmetic the \compswitch{-DLRSLONG} switch is set, with 
\compswitch{-DB128} also set
for 128-bit arithmetic. The \compswitch{-DSAFE} switch enables overflow checking and handling as
described in this section. The files \texttt{lrsarith.h} and 
\texttt{lrsarith.c} use these switches to ensure that the
correct arithmetic files are included. Lines 2 and 3 produce the binaries \fixedo and \fixedt.
For illustrative purposes we also compile without the \compswitch{-DSAFE} switch obtaining
\fixedon and \fixedtn in lines 4 and 5. In this case no overflow checking is performed.
With the \compswitch{-DMP} switch set the MP version is produced.
Finally by setting the \compswitch{-DGMP} switch we compile with the external GMP library
which must be preinstalled. 
For simplicity, we assume that the necessary files are in standard 
locations otherwise the locations need to be specified.
We ran \fixedo, \fixedt, \fixedm and \fixedg with the input $k=5$ getting the following output,
respectively: 

{\footnotesize
\begin{verbatim}
 5 25 625 390625 152587890625  overflow detected:halting
 5 25 625 390625 152587890625 2328364365386962890625  overflow detected:halting
 5 25 625 390625 152587890625 23283064365386962890625 542101086242752217003726400434970855712890625 
 5 25 625 390625 152587890625 23283064365386962890625 542101086242752217003726400434970855712890625 
\end{verbatim}
}
\fixedo can compute $5^{16}$ before overflowing and \fixedt can also compute $5^{32}$.
Both \fixedm and \fixedg can compute $5^{64}$. 
Running \fixedon and \fixedtn with no overflow protection
produces incorrect output.  This can observed by noting the three cases
where the last digit is not 5:

{\footnotesize
\begin{verbatim}
 5 25 625 390625 152587890625 3273344365508751233 7942358959831785217 
 5 25 625 390625 152587890625 23283064365386962890625 -30240059481632067979667719627811971327
\end{verbatim}
}
\subsection{Fixed precision performance}
\label{subsec:fixedperf}

We briefly compare performance of the various fixed arithmetic packages in
order to evaluate the overhead used by overflow checking.  While this can
also be seen in the more extended comparison done in Section~\ref{sec:ve},
here we use a simple code to further demonstrate the use of \lrsarith.

We consider a problem that uses a great deal of relatively
simple computations and little output.  The Collatz 
conjecture~\cite{collatz1,collatz2} is a famous open problem in number theory:
given a number $k$, we replace $k$ by $3k+1$ if $k$ is odd and by $k/2$ if
it is even.  The process continues in the same way and the question is
whether for every starting value, the sequence eventually reaches $1$.
There is a great deal of work~\cite{collatz1,collatz2} on the conjecture
and it has been~\cite{collatz2020}
verified\footnote{See also David Barina's page on the current status: \url{https://pcbarina.fit.vutbr.cz/}} for all $k<2^{68}$.

Our goal is to compare
arithmetic packages and not to computationally verify larger values, so we
avoid techniques such as huge lookup tables for simplicity.
We consider the Collatz sequence from $k$ as a path from
the integer $k$ to the value one. The set of such paths defines an infinite tree
with root $k=1$ and the conjecture holds if all integers appear in the tree.
One way to make the tree finite is to give a parameter
$\cmax$ and consider the tree of all paths to the root that do not contain
an integer $k$ 
greater than $\cmax$.
This finite tree can be generated from the root
without using memory to store its nodes using the reverse search procedure.

We wrote a code \coll\footnote{\coll is contained in \lrsarith-010.} to generate
the tree in this way. As long as $\cmax < 2^{63}/3 $ the program will not overflow
in 64-bit arithmetic. Table \ref{tab:Collatz} gives the running
times\footnote{\maif: 2x AMD EPYC7552 2.3GHz, 48 cores, 512 GB memory, 
               CentOS 7.6, gcc 4.8.5} when
the various arithmetic packages are compiled with \coll. 
The binary \collflint uses FLINT arithmetic (version 2.6.3) and the other binaries are labelled
in the same way as for \fixed above. 

\begin{table}[htbp]
\centering
\scalebox{0.90}{
\begin{tabular}[t]{|c|c||c|c|c|c|c|c|c|}
  \hline
$\cmax$&nodes&\colla&\collan&\collb&\collbn& \collgmp & \collflint&\collmp \\
\hline
$10^8$   &  39523168&  0.70&  0.60&   5.20&   4.63&   9.20&   7.24&   7.19 \\
$10^9$   & 395436300&  6.50&  5.79&  54.37&  48.40&  91.87&  72.26&  72.35 \\
$10^{10}$&3953296865& 82.28& 57.26& 570.26& 511.66& 923.77& 866.20& 864.84 \\
  \hline
\end{tabular}
}
\caption{Collatz tree generation: times in seconds Collatz tree 
(\maif)}
\label{tab:Collatz}
\end{table}

The values in Table~\ref{tab:Collatz} are small enough to avoid overflows,
but one can see that overflow checking in \lrslong results in only
minor overhead.
The relative performance of the packages will vary between machines,
mix of arithmetic operations, compilers and versions of extended
precision libraries; however, these
results give a sample of what can be obtained.
It is also worth noting that the multiprecision arithmetics are
generally focused more on larger operands.
In any case, for this computation on this machine, 
128-bit arithmetic is roughly eight times slower than 64-bit arithmetic.
FLINT arithmetic is somewhat faster than GMP (version 6.0)
but slower than the dedicated fixed precision arithmetics.
This is generally reasonable -- FLINT makes more effort to obtain
good performance on small operands, however \lrsarith fixed arithmetic
is essentially the same as using native arithmetic.

\section{Hybrid arithmetic}
\label{sec:hybrida}

The fixed arithmetic versions described in the previous section
are very simple to implement and easy to use, giving good
performance for inputs that do not require very long integers.
If overflow occurs the user can manually rerun the job with a
higher precision arithmetic package.
However, it is certainly more convenient to automatically
switch from one arithmetic to another with higher precision when
overflow could occur. In this section we describe how this can be done
using \lrsarith, while making only minor modifications to the original
user code.

\subsection{Combining fixed arithmetic packages}

We illustrate our approach using the example program of the previous section
that reads an integer $k$ and successively squares it.
The hybrid code is shown in Appendix \ref{app:hybridc}. 
Lines 9--29 contain the function \run that is essentially the function
\main from Appendix \ref{app:fixed}.
Instead of reading the integer $k$, it is passed to \run as a parameter.

The major change is the use of \setjmp to allow \lov to return control
to the program that calls \run. This is achieved via the variable 
\bufo defined on line
7. The main loop (lines 18--25) of \run is enclosed by a test on line 17.
If \lov (lines 31--33) is called by the arithmetic package
a long jump is performed to the statement immediately after the
end of the enclosed loop, i.e. line 26. This prints a warning
and returns to the routine that called \run.

In fact there are three routines that call \run, namely \runo, \runt, and \runmp (GMP or MP),
but only the first two can trigger a call to \lov.
The function \run itself
will need to be compiled three times, once for each package.
The arithmetic package used and the routine to call \run are selected at
compile time by compiler directives in the makefile and on lines 35--43,
respectively.
The problem now arises that since \run will be compiled three times the
three versions will need different names, a process known as
{\em name mangling}. 
In fact all of the routines 
that use \lrsmp will also need name mangling. The
particular scheme used in \lrsarith was suggested by David Bremner and
is illustrated in line 5. A unique suffix \suf defined in the
arithmetic header file used is added to the function \run.
These define lines will be needed for each user supplied
routine. The define for \lov is contained in \lrslong.h.

Listing 5 contains the \main program with its header file given in Listing 4.
It begins by reading the parameter $k$ and by
calling \runo which uses 64-bit arithmetic. If overflow occurs
a return code of 1 is triggered by \lov causing \texttt{main} to call
\runt using 128-bit arithmetic. If this in turn overflows a final call is
made to
\runmp which uses GMP\footnote{For simplicity we have omitted lines which
produce a hybrid executable with MP, but they are in the \lrsarith \make.}.
In the \make given in Listing 6, lines 4--6 compile the 
arithmetic libraries. Lines 7--9 compile \texttt{hybridlib.c} three times,
once with each arithmetic library. Finally line 11 combines everything into
a single executable \hybrid. 
Running \hybrid with $k=5$ produces:

{\footnotesize
\begin{verbatim}
 5 25 625 390625 152587890625  overflow detected:restarting
 5 25 625 390625 152587890625 2328364365386962890625  overflow detected:restarting
 5 25 625 390625 152587890625 23283064365386962890625 542101086242752217003726400434970855712890625 
\end{verbatim}
}

\subsection{Hybrid performance}
\label{subsec:hybridperf}

We continue the example of generating Collatz trees introduced in Section~\ref{subsec:fixedperf}. 
This time we use much larger values of $\cmax$ in order to create overflow conditions 
in both 64-bit and 128-bit arithmetic. As the trees now become extremely large, we
terminate the programs after $10^8$ nodes have been generated. The program
is deterministic so in each case the same nodes are traversed.
The program \coll
is the hybrid code combining \colla, \collb and \collgmp created in the same way
as \hybrid described above.

The results are shown in Table \ref{tab:Collatz2}.
On each line the final arithmetic used in \coll is shown in blue. It is interesting
to observe that for each of these three runs \collgmp runs in roughly the same time and 
that, for $\cmax=10^{32}$, this
is faster than the 128-bit arithmetic \collb (and hence \coll). 

\begin{table}[htbp]
\centering
\scalebox{0.99}{
\begin{tabular}[t]{|c||c||c|c|c||c||c|}
  \hline
$\cmax$&coll&\colla&\collb&\collgmp &\collmp & \collflint \\
 &(hybrid)&    & & & & (hybrid) \\
\hline
$10^{16}$& 
 \textcolor{red}{1.67}&\textcolor{blue}{1.70}& 18.60& 24.43& 22.93& 24.85\\
$10^{32}$&
 \textcolor{red}{36.21}&(o)& \textcolor{blue}{36.17}& 24.15& 34.10& 34.04\\
$10^{48}$&
 \textcolor{red}{25.59}&(o)&(o)&    \textcolor{blue}{24.66}& 45.83& 45.97\\
\hline
\end{tabular}
}
\begin{tablenotes}
\item[1]
~~~~~~~~~~~~~~~~~~~~(o)=possible overflow detected 
\end{tablenotes}
\caption{{Collatz tree generation: $10^8$ nodes, times in seconds  (\maif)}}
\label{tab:Collatz2}
\end{table}

The simple scheme described in this section
will be adequate in many applications but has a number of disadvantages.
First, memory allocated during the run for a given arithmetic
may not be freed after an overflow occurs, causing a memory leak.
More importantly, the simple scheme here restarts from the start of program
execution.  Forgetting the past can be helpful but this can hurt
performance if overflow occurs near the end of program execution.
It would be much better to not redo work that has already been done,
especially with slower arithmetic.

In applications such as \lrslib, these are serious issues. 
The definition of \texttt{lrs\_mp} depends on the arithmetic, so
offsets into structures containing \texttt{lrs\_mp} can change after
switching arithmetic.  Handling overflows appropriately is therefore
more complicated than simply forgetting the past and restarting,
as was described here.

The solution
used in \lrslib was to use a global variable to point to the allocated data structures.
This allows \lov to free the data after overflow is detected, before switching
to the next arithmetic. A further improvement
in \lrslib was to allow the program to restart at the point in the calculation
where the overflow was detected. Again this was done using a global variable
and also the ability of the original code to restart.
Fortunately this was already available.  In other programs, it may
be necessary to add periodic checkpoints and it can be helpful to separate
structures that contain \texttt{lrs\_mp} from those that don't.

\section{Vertex/facet enumeration problems}
\label{sec:ve}
When comparing~\cite{AJ15b} various codes for vertex/facet enumeration,
we noted that hybrid arithmetic could give a large speedup for certain inputs,
but that this was not available in \lrslib at the time.
We now turn our attention to the original motivation for developing the
hybrid version of \lrsarith: obtaining these speedups in \lrslib v.\ 7.

We begin by
introducing the basics of vertex and facet enumeration.  Then, we
explain the parallel hybrid version \mplrs, since this handles
overflow somewhat differently than hybrid \lrs.  
In Section \ref{subsec:setup}
we describe the experimental setup and in
 Section~\ref{subsec:arith}  we present a
comparison between the different arithmetic packages as used in \lrs and
\mplrs.

\subsection{Background}
\label{subsec:vebackground}
A convex polyhedron $P$ can be represented by either a list of vertices and extreme rays,
called a V-representation, or a list of its facet defining inequalities, called
an H-representation. 
The vertex enumeration problem is to convert
an H-representation to a V-representation. 
The computationally equivalent facet
enumeration problem performs the reverse transformation. 
For further background see Ziegler~\cite{Ziegler}. 
We will use the \lrs program in \lrslib to experiment with various instances
of these problems. A comparison with other codes including \mplrs, a parallel
version of \lrs, to solve these problems
can be found in \cite{AJ15b}.

The input is represented by an $m$ by $n$ matrix.
For a vertex enumeration problem this is a list of $m$ inequalities 
in $n-1$ variables whose intersection define $P$. 
A vertex (resp.\ extreme ray) is the intersection of a set of $n-1$ 
(resp.\ $n-2$) input inequalities, 
taken as equations, that satisfies
the remaining inequalities. A major difficulty is caused by degeneracy which
occurs when more than $n-1$ inequalities intersect at a vertex.
In this case \lrs generates multiple representations of the vertex, known as bases.
For a facet
enumeration problem the input is a list of the vertices of $P$ each beginning
with a 1 in column one\footnote{Extreme rays would be indicated by a zero in column one.}.
A facet is defined by a set of $n-1$ input vertices which span a hyperplane for which all
of the other input vertices lie on one side.
Here degeneracy is manifested when more than $n-1$ vertices lie on the hyperplane.
Again \lrs will generate the facet multiple times, each known as a basis.
When degeneracy occurs the vertex or facet is output when the index-wise
lexicographically minimum basis is found.

\subsection{Hybrid \mplrs}
\label{subsec:mplrs}

On detecting a possible overflow, the hybrid version of \lrs switches
to the next highest precision arithmetic package which it uses from
that point onwards.  The sequential implementation  was parallelized
as \mplrs~\cite{AJ15b} which dynamically partitions the work and
provides a series of subproblems to multiple \lrs workers.

If each of these workers was a hybrid \lrs process, it would start each
problem with the fastest arithmetic and switch to higher precision when
detecting a possible overflow.  This has a few drawbacks;
\begin{enumerate}
 \item Duplicate output: it's possible that when restarting after an
overflow, we reprint the most recent line of output.
 \item Performance: if overflow occurs, it usually occurs often in
 the run (ie not infrequently).  There is overhead in restarting and
switching arithmetic and so we would prefer to avoid frequent switches.
\end{enumerate}

In the hybrid version of \mplrs, each of the workers is a fixed
arithmetic process that returns to \mplrs when possible overflow is
detected.  At that point, \mplrs re-initializes the worker in question
using the next available arithmetic.  All output produced by the
subproblem is discarded and the worker restarts from the beginning (of the
subproblem).  The worker uses the new arithmetic from that point onwards.

This approach avoids duplicate output: by default, hybrid \mplrs holds
output produced by the worker if the arithmetic could overflow,
flushing it only when the job ends.  It also resolves the second problem;
each worker can only overflow and switch arithmetic twice in the overall
run -- the same as hybrid \lrs.  In addition, if overflow is a rare
event then it is possible for only some workers to overflow.  This
makes larger speedups possible comparable to hybrid \lrs.

There is overhead in two areas. First, when overflow is detected the
worker restarts its job.  This overhead is limited because jobs are
usually very small\footnote{See e.g., Figure 3(c,d) in \cite{AJ15b}.}
and workers can only overflow at most twice during the run.
Next, since \mplrs traverses different parts of the reverse search
tree in a different order compared to \lrs, it's possible for
overflow to occur earlier than it would in \lrs.  This could
hurt performance if \lrs overflows only at the end but could
also help performance by avoiding early \lrs overflows.

One complication is that for volume computation, \mplrs internally
uses the maximum precision arithmetic available.  This means that
\mplrs may not agree with the worker process on arithmetic.
Communication between the worker process and \mplrs therefore
avoids using \texttt{lrs\_mp} integers.

After checkpointing and restarting, hybrid \mplrs begins again 
with the fastest arithmetic available.  Checkpoint files are
compatible between the various arithmetic packages.

\subsection{Experimental setup}
\label{subsec:setup}
The polytopes we tested are described in Table \ref{tab:polytopes} and, except for
two new problems, were previously described and used in 
\cite{AJ15b}.
The problems range from non-degenerate
to highly degenerate polyhedra. This table includes the results of an
\lrs run
on each polytope as  \lrs
gives the number of bases in a symbolic perturbation of the polytope.
The new problems are:
\begin{itemize}
\item \cpseven is the cut polytope for 7 points which has input coefficients 0 or 1 and
is highly degenerate. 
\item \peight is related to the
holographic cone studied in physics and is an extension of the eight point cut polytope.  
Input coordinates are 0, 1 or -1 and it is the most degenerate problem studied.
\end{itemize}

We include a column labelled degeneracy which is the number of bases divided by the number of vertices (or facets) output, rounded to the
nearest integer. We have sorted the table in order of increasing degeneracy. The horizontal line separates the
non-degenerate from the degenerate problems.
The corresponding input files are available by following the download
link  at~\cite{lrs}.
Note that the input sizes are small, roughly comparable and much smaller than the output
sizes.

\begin{table}[htbp]
\centering
\begin{threeparttable}
\scalebox{0.8}{
\begin{tabular}[t]{|c||c|c|c|c||c|c||c|c|c|c|} 
  \hline
  Name &  \multicolumn{4}{|c||}{Input} & \multicolumn{2}{|c||}{Output} & \multicolumn{4}{|c|}{\lrs}\\
    & H/V & $m$ &$n$&size& V/H      & size & bases   &secs         &depth& degeneracy  \\
  \hline
\cthirty &V&30  &16& 4.7K& 341088   & 73.8M&319770   & 36          & 14    & 1  \\
\cforty  &V&40  &21& 12K & 40060020 & 15.6G&20030010 &7531         & 19    & 1  \\
\kmtwo   &H&44  &23& 4.8K& 4194304  &1.2G  & 4194304 & 234         & 22    & 1  \\
\permten &H&1023&11& 29K & 3628800  &127M  & 3628800 & 283        & 45    & 1  \\
\hline
\vffive  &V&500 &7 & 98K & 56669    &38M   & 202985  & 137         & 41    & 4  \\
\vfnine  &V&900 &7 & 20K & 55903    & 3.9M & 264385  & 23 & 45    & 5  \\
\mitseven&H&71  &61&9.5K & 3149579  &1.1G  & 57613364& 15474       & 20    & 18 \\
\fqfour  &H&48  &19& 2.1K& 119184   &8.7M  & 7843390 & 44          & 24    & 66 \\
\mitine  &H&729 &9 & 21K & 4862     & 196K & 1375608 & 132         & 101   & 283\\
\cpseven   &V& 64&22 & 3K & 116764     & 7.4M &308644212 & 4071& 41& 2643 \\
\bvseven &H&69  &57& 8.1K& 5040     & 867K &84707280 &1256         & 17    & 16807   \\
\peight &H& 154 & 92 & 36K& 4452  & 1.2M& 110640628    & 16152  & 63 & 24852 \\

  \hline
\end{tabular}
}
\caption{Polytopes tested and \lrs v.\ 7.1 times (\maif)}
\label{tab:polytopes}
\end{threeparttable}
\end{table}
\subsection{Comparison of arithmetic packages}
\label{subsec:arith}
In this section we give numerical results on using \lrslib to solve the vertex enumeration
problems described in the previous section with the various arithmetic packages in \lrsarith.
We test the following suite of codes, which apart from the exception noted below, are
available in \lrslib v.\ 7.1:

\begin{itemize}
\item
\lrsa :	64-bit fixed arithmetic with overflow checking.
\item
\lrsb :	128-bit fixed arithmetic with overflow checking. 
\item
\lrsgmp: 	GMP arithmetic (version 6.0). Comparable to default \lrs v.\ 6.2 and earlier.
\item
\lrs :	\lrsarith hybrid arithmetic. Starts with \lrsa and switches to \lrsb (if available) and finally \lrsgmp. Default version of lrs.
\item
\lrsMP\footnote{Available in v.\ 7.2 or on request.}:	As \lrs except in the final step internal MP arithmetic is used. 
\item
\lrsflint:	FLINT hybrid arithmetic (version 2.6.3).
\item
\mplrsgmp: 	Multi-core version of \lrsgmp, comparable to default \mplrs v.\ 6.2 and earlier.
\item
\mplrs:	Hybrid multi-core version of \lrs. Default version of \mplrs. 
\end{itemize}
Also available in  \lrslib v.\ 7.1 are the parallel versions \mplrsa, \mplrsb, and \mplrsflint
of the corresponding \lrs codes above. 

\begin{table}[htbp]
\centering
\scalebox{0.99}{
\begin{tabular}[t]{|c||c|c|c||c|c|c||c|c|} 
  \hline
  Name &  \multicolumn{3}{|c||}{Single Arithmetic } & \multicolumn{3}{|c||}{Hybrid Arithmetic}
       & \multicolumn{2}{|c|}{40 cores } \\
    &\lrsa&\lrsb& \lrsgmp & \textcolor{red}{\lrs}&\lrsMP &\lrsflint&\mplrsgmp &\mplrs \\
\hline
\cthirty &(o) &(o)&  \textcolor{blue}{36}   & \textcolor{red}{36}&295 & 42    & 2 & 3  \\
\cforty  &(o)&(o)& \textcolor{blue}{7707 }   &  \textcolor{red}{7581}& 97532 &8341 & 402 & 389  \\
\kmtwo   &(o)&(o)&  \textcolor{blue}{237}    &  \textcolor{red}{234}&384   & 187     & 8 & 8 \\
\permten &\textcolor{blue}{288}&538 & 2429  &\textcolor{red}{283}&284   & 1120      & 113 & 16 \\
\hline
\vffive  &(o)& (o)&  \textcolor{blue}{139}& \textcolor{red}{137}&1658 & 163    & 12 & 11 \\
\vfnine  &(o)& \textcolor{blue}{23}& 103&  \textcolor{red}{23}& 23  & 170    & 8 & 2 \\
\mitseven&(o)& (o)& \textcolor{blue}{15489}& \textcolor{red}{15474}&110467&26432& 731 & 723 \\
\fqfour  &\textcolor{blue}{44}&77& 265   & \textcolor{red}{44}&44   & 147  & 11 & 1 \\
\mitine  &(o)&\textcolor{blue}{134} & 563      & \textcolor{red}{132}& 131 & 270 & 27 & 7  \\
\cpseven & \textcolor{blue}{4105} &  6824&  25030 & \textcolor{red}{4071}&4039   & 14413  & 1141  & 191 \\
\bvseven &\textcolor{blue}{1276}  &2346  & 7981    & \textcolor{red}{1256}& 1252& 4874  & 357 & 65  \\
\peight  & \textcolor{blue}{16197}&23271&56519& \textcolor{red}{16152}&15970&45838  & 2395 & 661  \\
  \hline
\end{tabular}
}
\begin{tablenotes}
\item[1]
~~~~~~~~~~~(o)=possible overflow detected 
\end{tablenotes}
\caption{{Comparison of running times for various types of arithmetic, \lrslib v.\ 7.1 (\maif)}}
\label{tab:arithmetic}
\end{table}
The results of the tests are shown in Table \ref{tab:arithmetic}. 
The programs \lrsa and \lrsb
either produce the correct output or indicate that an overflow may occur (o) and terminate. 
The columns \lrs and \lrsMP give the running times for the hybrid versions.
When 64 or 128-bit arithmetic suffices these two running times are essentially the same. When extended precision
is required the internal MP arithmetic may be much slower than GMP, especially 
for \cforty, which requires 459 decimal
digits. The final two columns show the speedups obtained by using the 
multicore versions \mplrs and \mplrsgmp with 40 processors available.

In the table we show in blue which arithmetic version was in use when \lrs, shown in red,  terminated.
As to be expected \lrs performs roughly the same
as \lrsgmp when the integers become too large for fixed precision. Speedups of 
2--4
times are observed for the combinatorial problems which can be solved using only
64 or 128 bits. In general \lrsflint did not perform as well as \lrs
but it is usually faster than \lrsgmp 
on the combinatorial problems.
The approach used in \lrsarith hybrid arithmetic allows it to obtain essentially
native integer performance, but requires effort from the developer to handle
overflow.  FLINT is easier for the developer but did not achieve the same
performance when all values are small.
Another possible reason for the
lack of performance is that we did not use FLINT matrices as this would have required substantial
reprogramming of \lrs.

We can observe that 128-bit arithmetic runs roughly 1.5--2 times slower than 
64-bit
arithmetic, when 64-bit arithmetic suffices. 
Hence there is a strong incentive to start the computation with 64 bits
using 128 bits only when necessary. This latter outcome occurred for \mitine and \vfnine.

\section{Conclusion}
We introduced \lrsarith which is a small C library for performing fixed precision arithmetic
on integers and rationals
with overflow protection and allows hybrid arithmetic. It was developed as part of the \lrslib
polyhedral computation package. However, due to its small size and ease of use we decided
to release it as an independent package. The details of the method used and 
small examples
were given in the first part of the paper.

In the second part we gave computational results for some vertex enumeration problems.
The examples show the diversity of such problems and this is reflected in the
arithmetic precision required to solve them.
Many combinatorial polytopes involve calculations that can be completed without overflow
using 64 (or 128) bit integers. Using fixed arithmetic with overflow checking
or hybrid arithmetic, speedups of 2--4 times can be achieved and these
carry over to parallel implementations.  The new hybrid version described here
explains the performance improvements found in \lrslib version 7 relative to
previous versions or earlier comparisons~\cite{AJ15b}. 

\vspace{-0.08in}
\section*{Acknowledgements}
We thank David Bremner for 
many helpful discussions and in particular for the elegant implementation of name mangling.
This work was partially supported by JSPS Kakenhi Grants 
16H02785, 
18H05291, 
18K18027, 
20H00579, 
20H00595  
.
\vspace{-0.08in}
\bibliographystyle{plain}
\bibliography{lrsarith}

\appendix
\section{Single arithmetic code for repeated squaring}
\label{app:fixed}
\lstset
{ 
    language=C,
    basicstyle=\small,
    numberstyle=\tiny,
    numbers=left,
    stepnumber=1,
    showstringspaces=false,
    tabsize=1,
    breaklines=true,
    breakatwhitespace=false,
    numbersep=15pt,
}

\begin{lstlisting}[firstline=3,language=C,caption=fixed.c]
#include <stdio.h>
#include <stdlib.h>
#include "lrsarith.h"

FILE *lrs_ifp,*lrs_ofp;   

void lrs_overflow(int parm){
 fprintf(lrs_ofp,"  overflow detected:halting\n");
 exit (1);
}

int main (void){
 long i,k;
 lrs_mp a,b;
 lrs_mp_init(10,stdin,stdout);
 lrs_alloc_mp(a); lrs_alloc_mp(b);
 fscanf(lrs_ifp,"%ld",&k);
 itomp(k,b);
 pmp("",b);
 for(i=1;i<=6;i++)
  {
    copy(a,b);
    mulint(a,a,b);
    pmp("",b);
  }
 lrs_clear_mp(a); lrs_clear_mp(b);
 fprintf(lrs_ofp,"\n");
 return 0;
}
\end{lstlisting}

\begin{lstlisting}[firstline=16,lastline=22,language=C,caption=makefile]
fixed:	fixed.c lrsarith.c lrsarith.h lrslong.c lrslong.h lrsgmp.c lrsgmp.h lrsmp.c lrsmp.h
	$(CC) -DLRSLONG -DSAFE -o fixed1 lrsarith.c fixed.c
	$(CC) -DLRSLONG -DB128 -DSAFE -o fixed2 lrsarith.c fixed.c
	$(CC) -DLRSLONG -o fixed1n lrsarith.c fixed.c
	$(CC) -DLRSLONG -DB128 -o fixed2n lrsarith.c fixed.c
	$(CC) -DMP -o fixedmp lrsarith.c fixed.c
	$(CC) -DGMP -o fixedgmp lrsarith.c fixed.c -lgmp
\end{lstlisting}

\pagebreak

\section{Hybrid arithmetic code for repeated squaring}
\label{app:hybridc}

\begin{lstlisting}[firstline=4,language=C,caption=hybridlib.c]
#include <stdio.h>
#include <stdlib.h>
#include <setjmp.h>
#include "lrsarith.h"
#define run suf(run)

static jmp_buf buf1;   /* return location when overflowing */

int run(long k){
 long i;
 lrs_mp a,b;
 lrs_mp_init(0,stdin,stdout);
 lrs_alloc_mp(a); lrs_alloc_mp(b);
 itomp(k,b);
 pmp("",b);
 
 if (!setjmp(buf1)){ /* overflow test */
  for(i=1;i<=6;i++){
    copy(a,b);
    mulint(a,a,b);
    pmp("",b);
    }
  lrs_clear_mp(a); lrs_clear_mp(b);
  return 0;
 }                  
  printf("   overflow detected:restarting\n");
  lrs_clear_mp(a); lrs_clear_mp(b);
  return 1;
}

void lrs_overflow(int parm){
    longjmp(buf1,1);
}

#if defined(MA) && defined(LRSLONG)
#ifdef B128
int run2(long k){       /* 128 bit   */
#else
int run1(long k){       /* 64 bit    */
#endif
#else
int runmp(long k){      /* other arithmetic */
#endif

 return(run(k));
}
\end{lstlisting}

\pagebreak

\begin{lstlisting}[firstline=4,language=C,caption=hybrid.h]
#include <stdio.h>
#include <stdlib.h>

FILE *lrs_ifp; /* input file pointer */
FILE *lrs_ofp; /* output file pointer */

int run1(long k);
int run2(long k);
int runmp(long k);


\end{lstlisting}

\begin{lstlisting}[firstline=3,language=C,caption=hybrid.c]
int main (void) {
long k;

scanf("%ld",&k);

if(run1(k))          /* TRUE on 64 bit overflow  */
  if(run2(k))        /* TRUE on 128 bit overflow */
    runmp(k);

printf("\n");
return 0;
}
\end{lstlisting}

\begin{lstlisting}[firstline=3,lastline=13,language=C,caption=makefile]
HOBJ=lrslong1.o hybridlib1.o lrslong2.o lrsgmp.o hybridlib2.o hybridlibgmp.o

hybrid:	hybrid.c lrslong.c lrslong.h lrsgmp.c lrsgmp.h hybridlib.c hybrid.h lrsarith.c
	$(CC) -DMA -DSAFE -DLRSLONG -c -o lrslong1.o lrsarith.c
	$(CC) -DMA -DB128 -DSAFE -DLRSLONG -c -o lrslong2.o lrsarith.c
	$(CC) -DMA -DGMP -c -o lrsgmp.o lrsarith.c
	$(CC) -DMA -DSAFE -DLRSLONG -c -o hybridlib1.o hybridlib.c
	$(CC) -DMA -DB128 -DSAFE -DLRSLONG -c -o hybridlib2.o hybridlib.c
	$(CC) -DMA -DGMP -c -o hybridlibgmp.o hybridlib.c

	$(CC) -DMA -DLRSLONG -DSAFE -o hybrid ${HOBJ} hybrid.c -lgmp
\end{lstlisting}

\end{document}